\documentclass{article}
\usepackage{arxiv}
\usepackage{amsfonts}       
\usepackage{epsf}
\usepackage{epsfig}

\usepackage{amsmath}
\usepackage{xargs}
\usepackage[table]{xcolor}

\usepackage{booktabs}

\title{Gendered Networks and Communicability in Medieval Historical Narratives}

\author{
S~D~Prado, S~R~Dahmen\\
  Instituto de F\'{\i}sica\\
 Universidade Federal do Rio Grande do Sul\\
  Porto Alegre, 91501-970, Brazil\\
\And
A~L~C~Bazzan,\\
Instituto de Inform\'atica\\
 Universidade Federal do Rio Grande do Sul\\
  Porto Alegre, 91501-970, Brazil\\
\And
M~MacCarron\\
Department of Digital Humanities\\
University College Cork\\
Cork, T12 YN60, Ireland\\
\And
J~Hillner\\
Department of History\\
University of Sheffield\\
Sheffield, S3 7RA, UK \\
} 
\begin{document}
\maketitle
\begin{abstract}
One of the defining representations of women from medieval times is in the role of peaceweaver, that is, a woman was expected to 'weave' peace between warring men. The underlying assumption in scholarship on this topic is that female mediation lessens male violence. This stance can however be questioned since it may be the result of gender-based peace and diplomacy models that relegate women's roles to that of conduits between men. By analysing the concept of communicability and relevance of certain nodes in complex networks we show how our sources afford women more complex and nuanced social roles. As a case study we consider a historical narrative, namely Bede's {\it Ecclesiastical History of the English People}, which is a history of Britain from the first to eighth centuries AD and was immensely popular all over Europe in the Middle Ages.
\end{abstract}
\section{Introduction}

In the last few years we have witnessed an explosion in the use of networks as a quantitative tool with which one may analyse and quantify interpersonal relationships in human societies~\cite{Carrington2005,Borgatti2009}. Notwithstanding the fact that applications of networks in the Social Sciences dates back to more than 5 decades~\cite{Wasserman1994}, it was only in the last few years that there has been renewed interest in the subject mainly due to the increase in computational capability at our disposal. More recently, historians have been taking heed of these new methodologies in the hope of gaining new insights into their subject matter~\cite{Bearman2002,Gould2003,LemercierZalc2019}. This has been seconded by more detailed studies of character networks in narratives, fictional or not~\cite{Kenna2017,Yose2018,Dahmen2016,Bost2019}. The narration of historical events entailed the description and construction of complex relationships between large numbers of actors. Such events may span days, months, years or even centuries, making them also ideal proving grounds for the study of networks that evolve in time~\cite{Holme2015}. However,  as a consequence of our inability to deal in a concise and coherent fashion with such a vast amount of information, historians' interpretations tend to concentrate on a few key figures and events in the hope that some major threads will help them better understand the past. This does not mean that our sources support the epistemological viewpoint purported by the so-called 'great man theory' of Thomas Carlyle~\cite{carlyle1841}, namely that it is the select few that define the course of history. This stance
however has shaped the way we perceive history and has led, among other things, to the portraying of women as subsidiary to main historical events. Gendered assumptions structuring present society have deep roots in the past, and the question remains whether past societies afforded women more expansive roles in community and network building than we often realise. Seen from this perspective, network theory can be of great value as it unveils complex hidden structures which can be determinant to the way lesser known historical figures played an important role in historical events.

The early middle ages, for our purposes specifically the period from ca. 330 to 735 AD, is a particularly fruitful era for studying the role of women as connectors of men. It was a period 
2
 that experienced rapid changes in political authority and was marred by warfare, feud and religious conflict. Read in an analogue way, texts from this period have often been interpreted as reducing women's social role to that of either sowing conflict between men or reconciling them. This is demonstrated, for example, by historians' usage of the term {\it peaceweavers}, a direct borrowing from the Old English kenning
{\it freothuwebbe} for royal women who married outside their kingdoms~\cite{Rosenthal1966,Nelson1978}. They argue that, in society at least, the primary social expectation of women was to connect men. These approaches overlook, however, that it was not only their gender that must have
determined the extent of early medieval women's power, but also their social relationships and their
changing positions within various social networks. This view of women's primary role as intermediaries between men has been challenged in the last few decades~\cite{Rooney2002,Meth2010,Ramnarain2015}.
With this in mind we take here a more detailed look at a network of historical figures described in Bede's {\it Historia Ecclesiastica Gentis Anglorum} (henceforth called {\it HE}) in order to assess the role women played in early medieval history~\cite{BedeHE}. Bede's {\it HE} offers an unusually comprehensive and coherent picture of events in Britain during this period. Bede's {\it HE} also allows us to explore different kinds of networks, as the links connecting people
are categorised according to 21 different types of relationships (we describe these in more detail in the next section). 

In this paper we analyse the communicability of nodes, {\it i.e.} their ability to send or receive information~\cite{Estrada2008,Grindrod2011,Estrada2012}. We compare temporal measures with aggregate ones and discuss the importance of characters according to the definitions of~\cite{Shetty2005,Qiao2017,XinboAi2017}.
Our results show that some women were indeed very important within the network structure of the narrative and this importance is unrelated to the powerful men to which they were connected. The constitutive role of certain women in this historical character network would not emerge without the techniques applied here.

This paper is organized as follows: in the next section we briefly discuss Bede's work in its historical context and the reason for choosing it as a case study. This is followed by a section on communicability for aggregate and temporal networks and the definition of node relevance. We then present our results, with particular emphasis on women and discuss those from a historical viewpoint. We conclude with some perspectives for future work and discussion on the use of networks in history.

%
%
\section{Bede's Ecclesiastical History of the English People and its Network}

The Venerable Bede ({\it c.} 672/3 -- 735) was a monk in Northumbria and one of the foremost scholar of Western Europe of his time, writing extensively on theology, history, mathematics, natural phenomena and time reckoning~\cite{Bede2009,Bede2010,MacCarron2019}.

However, in spite of the extensive range of his literary production, he has gone down in history as the 'Father of English History' for a good reason: his is the most detailed account of the early history of Britain, and the main source of information we have from that period. Bede was part of extensive intellectual and ecclesiastical networks. He was in close contact with the archbishopric of Canterbury and other monasteries and churches throughout Britain many of whom assisted in his efforts to learn about the history of his people. Thanks to the efforts of the founding abbot of his monastery, Benedict Biscop (d. 690), who visited Rome five times and always returned with books and other ecclesiastical resources, Bede had access to one of the best stocked libraries anywhere in Europe at this time. The depth of his learning is apparent in his historical, theological and scientific writings. In the {\it HE}, Bede endeavoured to reveal his own people's role in the unfolding of salvation history by recounting the story of their conversion to Christianity and as they became members of the universal church.  His work encompasses approximately $800$ years, from the Roman conquest of Britain by Julius Caesar in $55$ {\it BC} to the first half of the 8{\it th} century.

Bede's {\it HE} is divided into five books which are to a certain extent chronologically ordered. These books are further divided into chapters, ranging from 20 to 34 depending on the book. Book 1 opens with a description of Britain's geographical position and covers the longest period of time running from Julius Caesar's first invasion in the first century BC up to the late sixth century and the arrival in Kent of Christian missionaries sent from Rome. Books 2 to 5 recount the gradual movement of Christianity from Kent to the northern kingdom of Northumbria, including the significant setbacks experience by the missionaries, and by the conclusion of the book Bede indicates that the whole island has accepted Christianity. Books 3 and 4 chart the movement from early Christianity to the mature establishment of the Church including the proliferation of monasteries in the second half of the seventh century. Historians have long noticed the importance of books 3 and 4 for the {\it HE}'s overarching narrative and our examination has confirmed that in terms of the number of people and connections, books 3 and 4 dominate the narrative.

\subsection{Network and data gathering}

In order to compare the properties of a range of different sources from the fourth to the eighth centuries, in this
project we have developed a data model that classifies the types of relationship we encounter in our sources into 21 different categories of relationship. This model offers depth of analysis as we can distinguish the relationship between spouses, from those of family members, clerics in a monastic community, or military combatants to name but some. However, it is not a number large enough to render some of the networks too sparse thereby making the analysis devoid of meaning. This means, for instance, that one can construct 21 different networks for any given period of Bede's
narrative. Some can be very sparse while others are not, depending on the category or combination of categories one
chooses. The categories can be further grouped into two broad sets, namely friendly and hostile relations. The
friendly ones are: kinship, marriage, concubinage, domestic relations, fosterage, friendship, church meetings, monastic family\footnote{Monastic family entails connections between people who were part of the same church community, but not necessarily in the same place.}, spiritual kinship, patronage, political connections, physical diplomacy, letter recipients and textual transmission. Of these relationships, three are directed: letter recipients, textual transmission\footnote{Textual transmission refers to a person who has read a work by someone they have never met.} and patronage. Hostilities can be of a
military nature, political, religious, domestic or gender-based violence, all undirected. Two other categories, which
can be friendly or hostile are {\it post mortem}, which describes interactions between people and the deceased as well as supernatural, when angels, spirits, demons and so on intervene in daily affairs. A modern reader might dismiss  these last two categories as irrelevant, but for the people living in the Middle Ages these interventions were real and were an inherent part of the narrative, so for the sake of completeness we have recorded them. However, as these are separate relationship categories we can also remove them from our analysis of the network should we choose to.

Whenever two or more persons are described by Bede in some sort of interaction, an edge between them is recorded under the category that best represents that interaction. That is, we connect characters if we are explicitly told that they are connected, such as when two characters are presented as brothers, or implicitly when two characters are presented as part of the same enterprise at the same time, for example, members of the same missionary team. This is done manually as this is the only way that we can record networks with such depth in relationship categories. Moreover, many of the characters are unnamed or they are a collective group and any automated reading method would not be able to single them out (see Fig. ~\ref{fig:fig2} for an example network; note that some of the blue circles are unnamed women). We also work from the original Latin. Edges are weighted, that is we associate with them a value $\ge 1$ which accounts for the number of times the corresponding nodes interacted throughout the narrative. When categories are combined, edge weights are added. 


As we shall argue in more detail below, when studying the ability of nodes to communicate we must differentiate between their ability to send or receive information. For directed networks this is a given, but for undirected ones this difference arises as an effect of temporal ordering of links~\cite{Estrada2008}. Since we were interested in this time-generated asymmetry, we studied only the network which excludes directed links, namely letter recipients, textual transmission and patronage. Moreover we did not consider {\it post-mortem} and supernatural as these can have a mix of directed and undirected links within the same category. Moreover, Bede has no instance of gender-based violence in his narrative, so we effectively worked with a network composed of 15 undirected categories. In this case the network has $473$ characters, of which $62$ are women. The number of links is $2340$. Below we present a table with the number of people $P$ and links $L$ for each book:

\begin{center}
\begin{tabular}{lcccc}
\toprule
 && P & L & {\small\% of links} \\
\midrule
Book $1$ && 87    & 276   & 12\\ \\ 
Book $2$ && 89    & 456   & 19\\ \\ 
Book $3$ && 143   & 712   & 31\\ \\ 
Book $4$ && 144   & 572   & 24\\ \\ 
Book $5$ && 117   & 324   & 14\\ \\ 
total    && 473   & 2340  & 100\\  
\bottomrule
\end{tabular}
\label{tab:table1}
\end{center}
\vskip 0.5cm
{\small Table 1: Number of people/vertices $P$ and undirected links/edges $L$ of Bede's {\it HE}. The last column is the percentage of links of each book relative to the total number of links.}

%
Even thought there is one more character in Book 4 as compared to Book 3, the latter has a considerably larger number of links and consequently more weight than all other books. The effect of this dominance of book $3$ will be discussed in the section where we present our results.

\section{Methods}

In this paper we use two measures to assess the relevance of characters in the narrative of Bede: communicability and importance\footnote{Henceforth we use the terms relevance and importance interchangeably.}. We discuss both measures in details in what follows.

\subsection{Communicability}

The communicability between a pair of nodes is usually defined as the shortest path connecting them. It owes its name to the idea that even when any two nodes are not directly connected, a signal starting from one can reach the other through intervening nodes. In this context, communicability plays a role in epidemic spreading, signal transmission and the submission/adjustment to social normal, to name a few examples~\cite{Estrada2012}. Intuitively it is also clear that in certain situations the communication between any
two given nodes might not happen through the shortest possible path between them but by means of longer ones. Thus one
may generalize the idea of 'path' between any two nodes $i$ and $j$ to that of a 'walk'~\cite{Estrada2008}.  A path of length $k$ is a sequence of $k$ {\it different} edges $e_1, e_2, \cdots e_{k-1}, e_k$ connecting $i$ to $j$. On the other hand, a {\it walk} of length $k$ between $i$ and $j$ is a sequence of not necessarily different edges connecting both nodes, {\it i.e.} edges can be traversed more than once on the way from $i$ to $j$. A path implies an edge being crossed only once, whereas in a walk we are allowed to pass the same edge again and again as long as it is undirected. Directed edges can be traversed only once in a specific direction. This idea can be made more precise as follows.

A network is a set of vertices $\{V \}$ and edges $\{E \}$ that can be depicted as a graph $G=(V,E)$. Let  $|V|= P$ be the number of vertices and $|E|= L$ the number of edges. For this graph one may define the adjacency matrix $A(G)=\mathbf A$ whose elements are given by:
\begin{equation}
A_{ij}=
\begin{cases}
1&\text{if } i\, \text{and}\, j\,\,\, \text{are connected} \\
0&\text{if } 0 \,\,\, \text{otherwise.}
\end{cases}
\end{equation}
It is a known fact in the field of algebraic graph theory that the $(i,j)-$entry of the {\it k}-th power of the adjacency matrix $(\mathbf{A}^k)_{ij}$ is equal to the number of walks of length $k$ between nodes $i$ and
$j$~ \cite{Estrada2008,Biggs2003}. Therefore one may define the communicability $C_{ij}$ between nodes $i$ and $j$ as the sum of all walks between them:
\begin{equation}
C_{ij} = \mathbf{A}_{ij} +(\mathbf{A}^2)_{ij} + (\mathbf{A}^3)_{ij} + \cdots
\label{eq:commsimple}
\end{equation}
Undirected links can make this sum diverge. In order to circumvent this problem we can modify it by including a 'damping' factor $\alpha < 1 $ so that the sum
converges for a properly chosen value of $\alpha$~\cite{Fenu2015}:

\begin{equation}
\mathbf{C} = \alpha \mathbf{A}  + \alpha^2 \mathbf{A}^2 + \alpha^3  \mathbf{A}^3 + \cdots = (\mathbf{I} - \alpha \mathbf{A})^{-1} = Res\,(\mathbf{A})
\label{eq:comm}
\end{equation}
where $Res (\mathbf{A})$ is the resolvent of $\mathbf{A}$ and $\mathbf{I}$ the identity matrix. It can be shown that if $\alpha=  1/|\lambda_{max}|$ then convergence of (\ref{eq:comm}) is guaranteed. $\lambda_{max}$ is the largest eigenvalue of $\mathbf{A}$ and $|\lambda_{max}|$ is the so-called spectral radius of $\mathbf{A}$.

Definition (\ref{eq:commsimple}) can be applied to adjacency matrices representing directed or undirected networks. For the context of the present work its most advantageous property is that it can be easily extended to treat temporal networks~\cite{Estrada2008,Fenu2015}. For the sake of completeness we briefly discuss this extension before we present our results.

Networks can change in time~\cite{Holme2012}. This is also true for character networks described in historical sources, because characters and the events they participate in are arranged sequentially and therefore are intrinsically dynamic. The natural way of thinking of a network represented by an adjacency with entries $A_{ij}$ which
has $M$ different realizations over a given period of time is to extend it to a rank-$3$ tensor of the type $A_{ijk}$
where $k=1,2,3,\cdots M$. In order to handle this rank-$3$ tensor by the known methods of network analysis one usually
resorts to  the 'flattening' of a rank-$3$ to a rank-$2$ tensor. One such flattening method was introduced by Taylor and coworkers~\cite{Taylor2017} which consists of combining the adjacency matrices of different times into one large supra-adjacency matrix

\begin{equation}
\label{eq:supermatrix}
\mathbb{A} =
  \begin{bmatrix}
    \varepsilon\mathbf{A}^{(1)} & \mathbf{I} & 0 & \cdots  \\
     \mathbf{I} & \varepsilon\mathbf{A}^{(2)} & \mathbf{I} & \ddots  \\
     0 & \mathbf{I}  & \varepsilon\mathbf{A}^{(3)} & \ddots \\
    \vdots & \ddots & \ddots & \ddots 
  \end{bmatrix}
\end{equation}
Each block diagonal $(P\times P)$-matrix $\mathbf{A}^{(t)}$ represents the adjacency matrix $\mathbf{A}(t)$ at a given time
layer $t$ ($t=1,2,\cdots M$)~\cite{Taylor2017}. This $PM \times PM$ matrix couples each matrix at a given time $t$ with its realization at time $t+1$ via the off-diagonal identity matrices. The quantity $\varepsilon$ plays the role of a coupling strength between different times. Even though this method has been successfully used in the study of narrative networks~\cite{Dahmen2016}, the lower diagonal is a mathematical condition imposed to guarantee the positivity of the matrix  eigenvalues, as they represent centralities (Perron-Frobenius Theorem). However it implies backward-causation in time, which is physically unacceptable. When it comes to communicability which, simply put, can be seen as the ability of a node to send or receive information, this leads to the unphysical situation where a node can send messages to the past (or receive it from the future). The definition of communicability in Eq. (~\ref{eq:comm}) remedies this situation if one removes the lower diagonal,  leading to  meaningful results where back causation is forbidden. Moreover it can also be easily extended
to deal with time-varying adjacency matrix. One may define the temporal version of Eq. (~\ref{eq:comm}) as
\begin{equation}
\mathbf{C} = (\mathbf{I} - \alpha \mathbf{A^{(1)}})^{-1} \, (\mathbf{I} - \alpha \mathbf{A^{(2)}})^{-1}
\, \cdots \, (\mathbf{I} - \alpha \mathbf{A^{(M)}})^{-1}
\label{eq:commtime}
\end{equation}
$\mathbf{A^{(t)}}$ represents the adjacency matrix at a given time $t$ ($t=1,2,\cdots M$). It is important to note that
in this case, to guarantee the convergence of the product, the parameter $\alpha$ is defined as before but using for its definition the largest eigenvalue among the $M$ highest eigenvalues of each adjacency matrix $\mathbf{A^{(t)}}$, that is $\alpha = max\{1/\rho^{(1)}, 1/\rho^{(2)},\cdots, 1/\rho^{(M)}\}$. Defined this way, the communicability can be viewed as the number of walks across space and time using a time-ordered sequence of adjacency matrices $\{ \mathbf{A^{(t)}} \}_{t=1}^M$.

From this one may define two different instances of communicability, a so-called {\it in}--communicability (receive-communicability) and an out-communicability (broadcast-communicability). It is given by the sum of
the entries of a row or column of $\mathbf{C}$ respectively~\cite{Fenu2015}:

\begin{equation}
    C^{out} = \mathbf{C}^T\,\mathbf{1}\,\,\,\,\,\,\,\, C^{in} = \mathbf{C}\,\,\mathbf{1}
    \label{eq:cout}
\end{equation}
In this equation $\mathbf{1}$ is an all-ones vector and $\mathbf{C}^T$ is the transpose of $\mathbf{C}$.

\subsection{Importance}
\label{sec:sec32}
The entropy of a network characterizes the amount of information encoded in the network structure. So, by removing one node and recalculating the entropy of the network, we can rank the changes in entropy, or network structure, from its highest to its lowest value and thus see how the removal of a given node affects it. Nodes which cause the largest entropy change are supposedly the most influential. We compute the entropy of a given node $j$ as \cite{Qiao2017,XinboAi2017}:

\begin{equation}
S_j=-\rho_j \log{\rho_j}
\label{eq:entropyindividuum}
\end{equation}
where $\rho_j$ is a probability density defined via the communicability $C_j$ of node $j$:
\begin{equation}
\rho_j=\frac{C_j}{\sum_{j=1}^{P} C_j}
\label{eq:densityindividuum}
\end{equation}
The importance $I_j$ of node $j$ can be defined as the difference between the overall entropy, when all nodes are considered, and the
same entropy when node $j$ is left out
\begin{equation}
 I_j = \frac{S-\overline{S}_j}{S}
\label{eq:importance} 
\end{equation}
with $S=\sum_{j=1}^{P}S_j$ and $\overline{S}_j=\sum_{k\neq j} S_k$.

%
%
In the following section these concepts are employed on the data described in Section 2.

\section{Data Analysis}

\begin{figure}[h]
\begin{center}
\includegraphics[scale=0.5,angle=270]{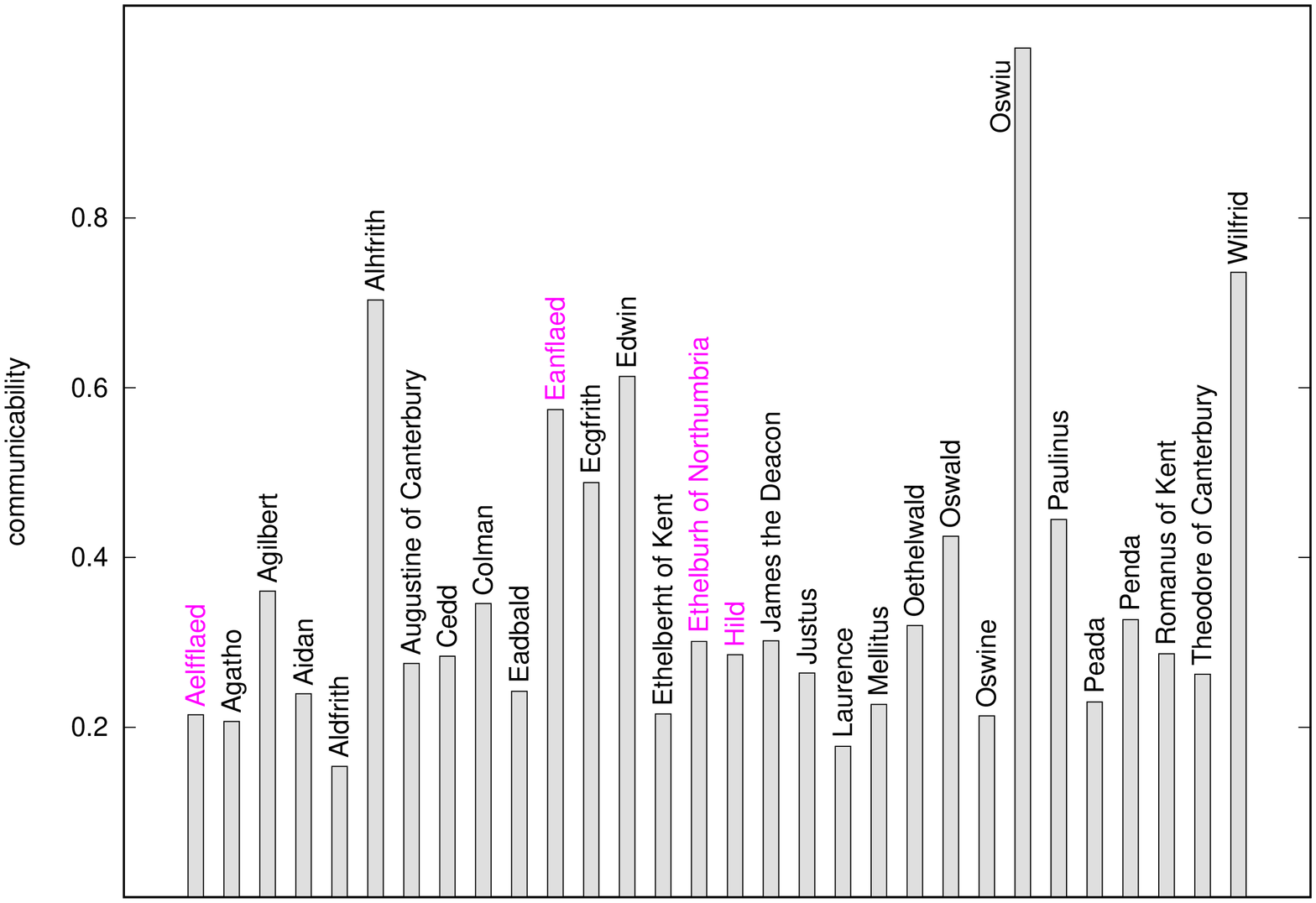}
\includegraphics[scale=0.5,angle=270]{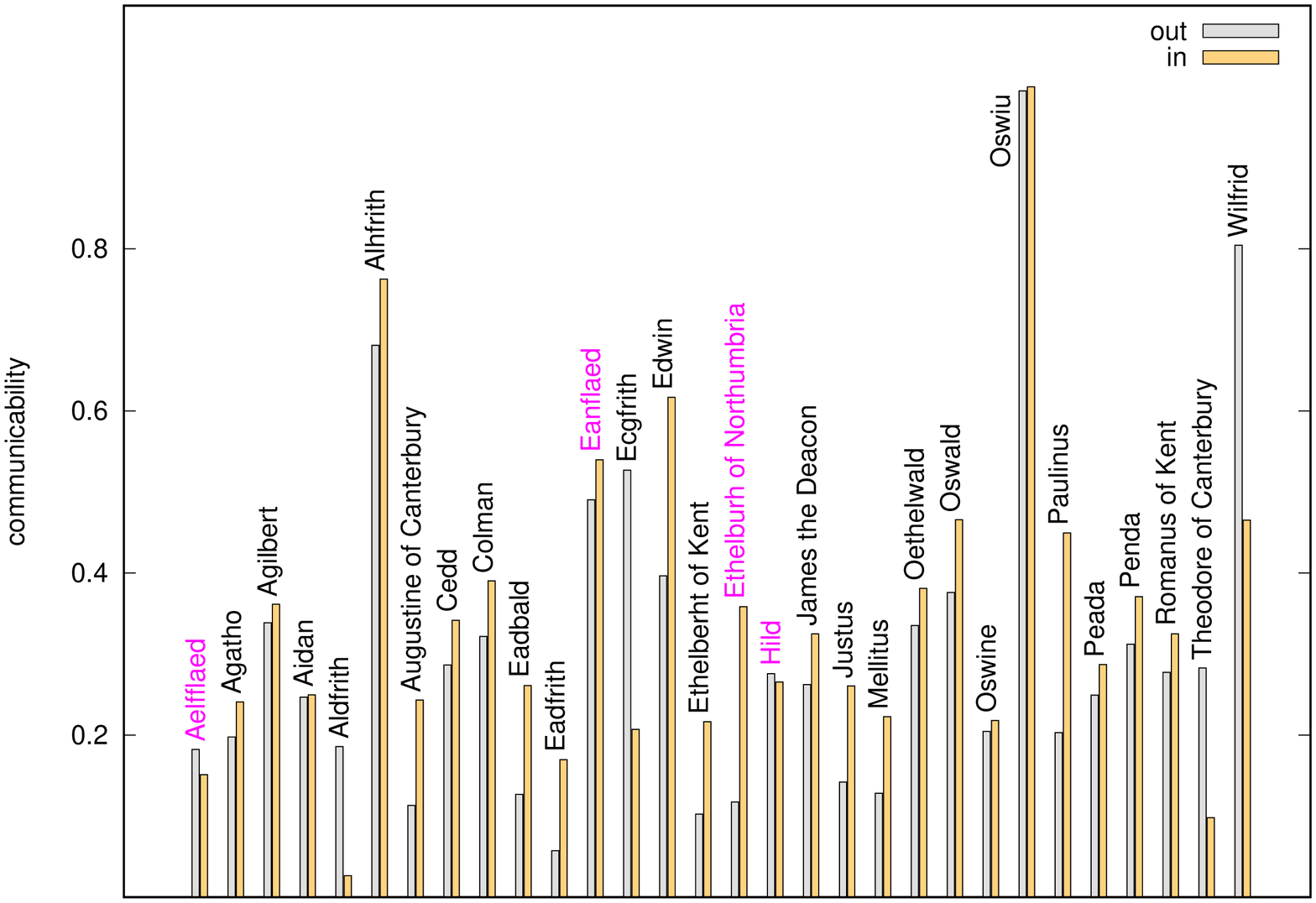}
\caption{{\small Top tier characters in communicability values for the aggregate (upper panel) and temporal networks (lower panel). }}
\label{fig:fig1}
\end{center}
\end{figure}

We are interested in determining which women, if any, play an important role in  Bede's {\it HE} network. In order to find this, we first calculated the communicability in the aggregate and temporal approaches. In the aggregate one, for undirected links, there is no distinction between receiver/broadcaster (in/out) communicability. However, in the  temporal case,  the same undirected links may give rise to two different values of communicability due to links being time-ordered.  The results are depicted in Fig. (\ref{fig:fig1}). For the sake of clarity we
show only the 30 highest-ranking characters (highest communicability value  normalized to 1.0).

First of all one can see some remarkable differences between in- and out-communicabilities (lower panel), the largest difference being that of Saint Wilfrid: as a founder and  ruler of many monasteries, educator and an artistic patron his influence was wide felt in Northern Britain. As a Northumbrian noble, he entered religious life at a young age and spent part of his formative years studying in Gaul and Rome. He was regarded as the spokesman for Roman rules of the church, as opposed to the Irish monks at Iona and played an important role in the Synod of Whitby, held in 664, where these questions were discussed. Second to Wilfrid is  Theodore of Canterbury, an archbishop and church reformer who appointed many bishops to different sees in Britain. He founded a school in Canterbury which gave birth to the so-called 'golden age' of scholarship in Northern Britain, hence his relevance. Other notable differences are that of Ecgfrith and Aldfrith, both sons of King Oswiu, who later became kings of Northumbria.

Another feature to notice is that the relative height of the bars and the order of most of the characters, from lowest to highest communicability, remain practically unchanged  when one compares both results. One would expect measures obtained from temporal networks to differ significantly from those obtained in aggregate ones, particularly if this measure is the communicability. This is easy to understand: if links are not directed and not time ordered, the network is just a structure of nodes connected by links which can be traversed in both directions. However, for the case of time-ordered links, even when they are not directed, there will appear an asymmetry in the sending or receiving of information: once a link from, say node $i$ to node $j$ has been traversed at a time $t_1$, we cannot retrace our step at a later time $t_2$ because the link might not exist anymore. Any message sent from $i$ to $j$ previously will still be able to propagate through the network but will not
be able to trace its path back, as the link disappeared. So in temporal networks the distinction between the two types of communicability comes naturally and is a consequence of time-ordering \cite{Fenu2015}.

It goes without saying that networks representing historical characters are by construct temporal ones. However, looking at Eq. (\ref{eq:commtime}) it can also be seen that the parameter $\alpha$ is selected from a given realization $\mathbf{A}^{(t)}$ which has the largest eigenvalue among all other adjacency matrix. This leads, in some cases, to one particular time frame dominating over others, in which case one
will not see much difference in the aggregate or temporal version of the network.

This is exactly the case for Bede and can be understood in terms of the way his narrative was written: each book and chapter of Bede's {\it History} encompass a different number of years, as some events he reports took place hundreds of years before his time while he was contemporary to some others. So the question is how to break them into more or less self-contained historical periods. In the case of Bede our choice was to use the books he divided his work into as a natural division of time, as they are temporally ordered in spite of not representing the same length of time. As Book $3$ is larger than $4$, it will dominate the narrative and this explains why there will be little difference between the temporal and aggregate versions of the network. In the temporal case, the parameter $\alpha$ comes from the adjacency matrix of book $3$. In the aggregate case, the portion of the entire network which come from book $3$ is denser.

Notwithstanding these important differences, given that we are interested in gendered networks and that the same women (names in magenta in Fig.~\ref{fig:fig1}) appear in both versions with the same relative rank order, we will concentrate in the forthcoming discussion on the aggregate version only. Although this is not the appropriate choice in general, since for some networks temporal order may yield rather profoundly different results, it is computationally simpler and justified in our particular case. 

\begin{figure}
\begin{center}
\includegraphics[scale=0.5,angle=0]{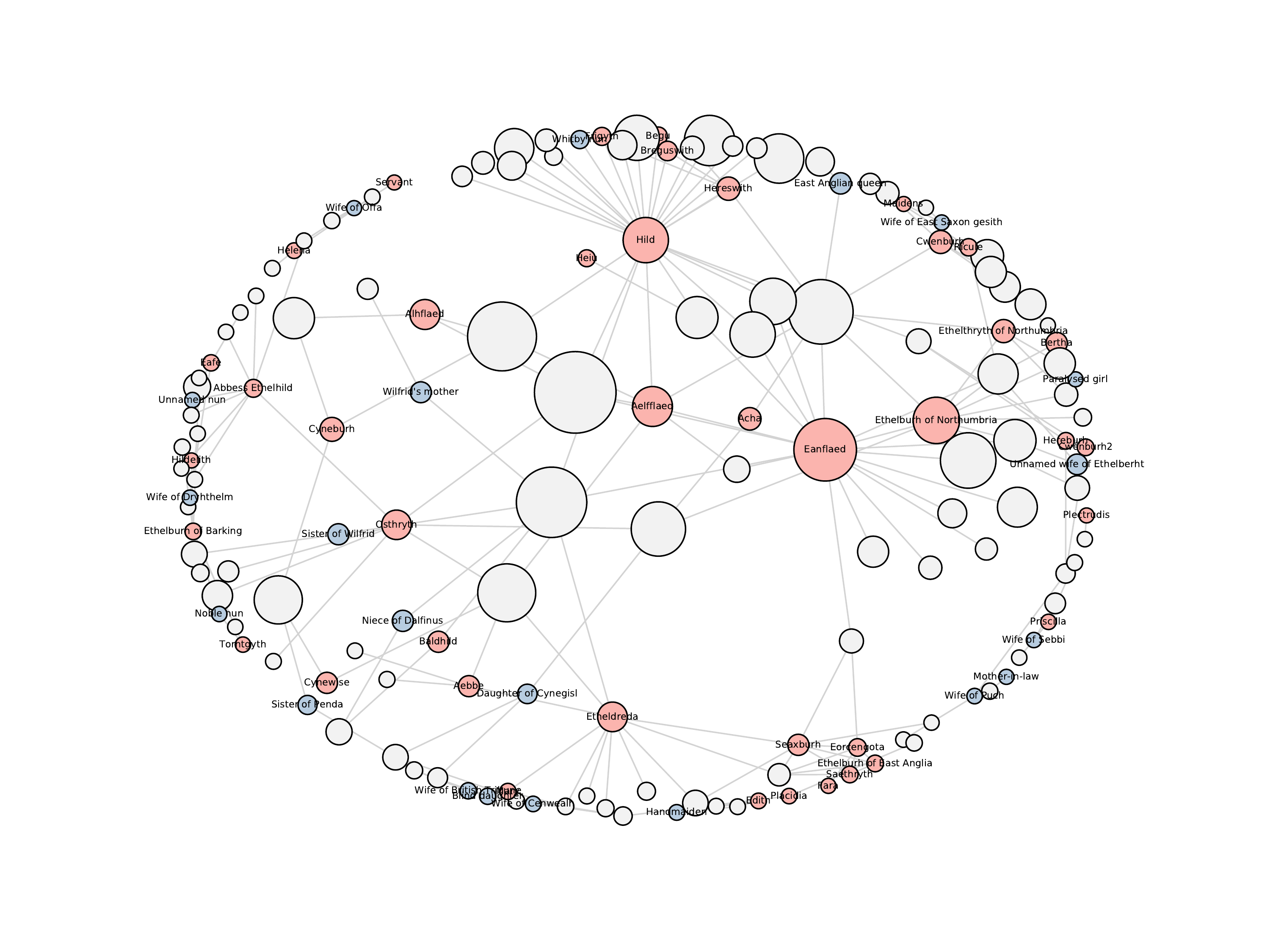}
\caption{{\small The network of characters according to their communicability. The size of each circle is proportional do the communicability of a given node. Women are depicted as salmon-coloured. Blue circles represent women who are unnamed in the text. Men are depicted as grey circles. Note how Queen Eanflaed is directly connected to the largest circle which represents the character with largest communicability, King Oswiu, her husband. For the sake of clarity, we did not encode
the weights of the edges in width of the lines that represent them.}}
\label{fig:fig2}
\end{center}
\end{figure}

A surprising result is the appearance of four women among the main characters, one of them -- Eanflaed -- having the fifth highest communicability value. Based on the historical context of the narrative, the appearance of Eanflaed raises a red flag: she was a Queen to King Oswiu of Northumbria, who is the historical persona with the highest communicability and it could mean that her result might be a reflection of his connections. This becomes clear when one examines the network depicted in Fig. (\ref{fig:fig2}).

In order to check if a woman's high communicability is a reflection of her being attached to a high-ranking men, we verified the relevance of characters by assessing how the topology of the network is changed by their removal. This can be done in different ways but we opted follow the definition of entropy~\cite{Shetty2005} as this measure is very sensitive to changes in the overall structure of the network \cite{Borgatti2006,Arroyo2010}. This is discussed in what follows (see section~\ref{sec:sec32} for the appropriate definitions).

\begin{figure}
\begin{center}
\includegraphics[scale=0.5,angle=270]{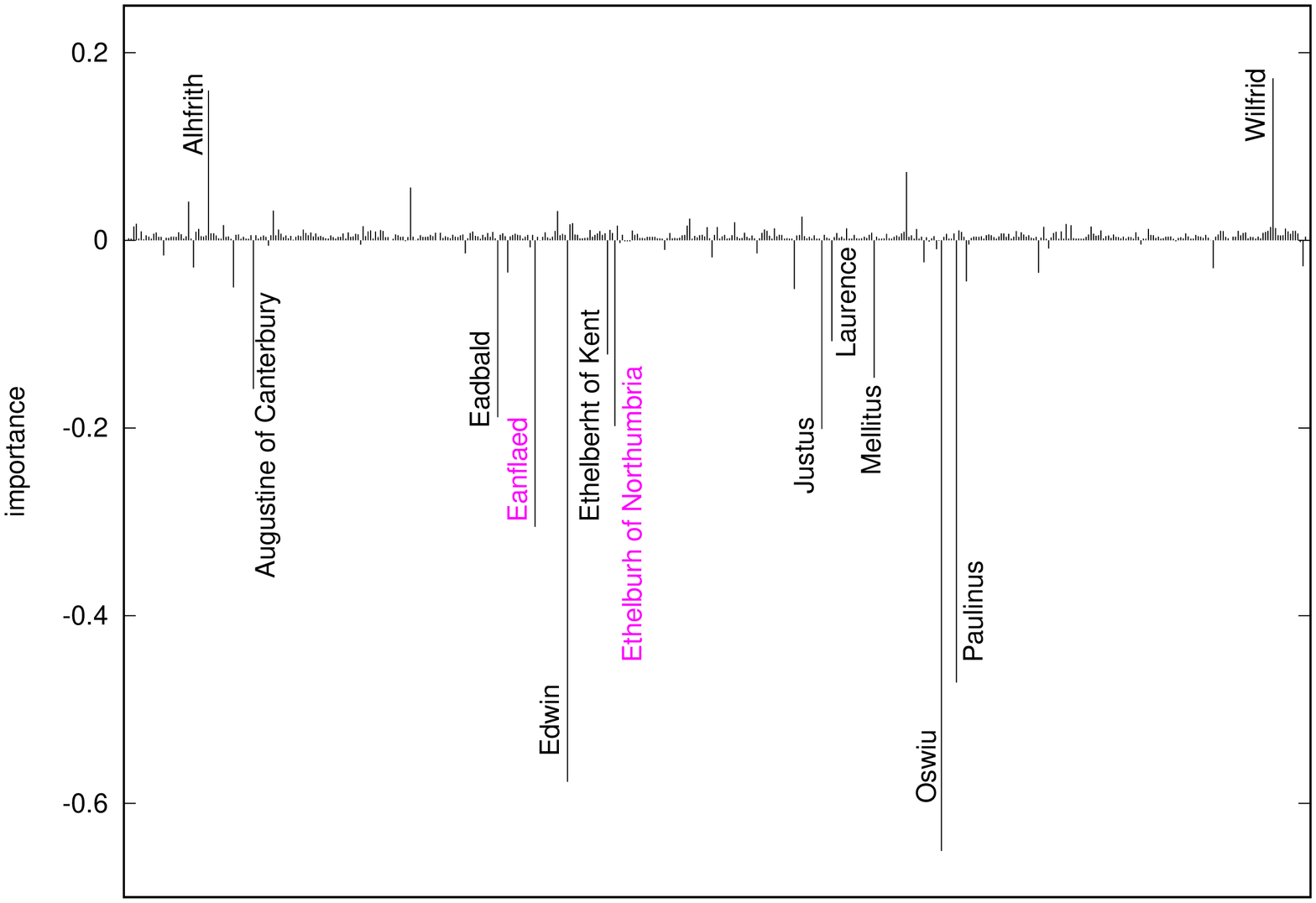}
\includegraphics[scale=0.5,angle=270]{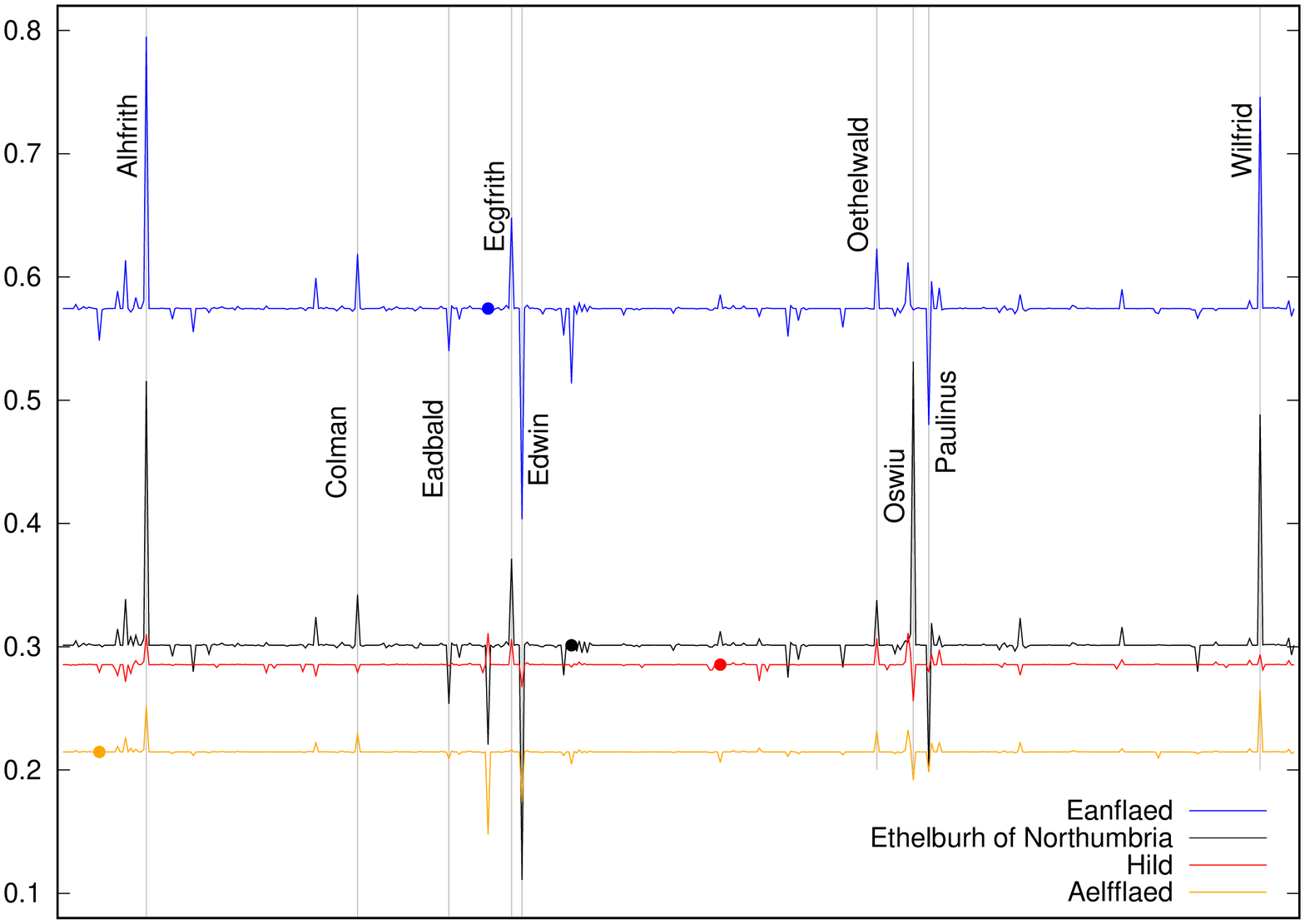}
\caption{{\small Upper panel: the importance of all characters given by the relative deviation of the network's entropy as defined by Eq. (\ref{eq:importance}). Lower panel: the variation of the communicability of the four most relevant women when characters are removed from the network. The removed characters with highest variation are represented by vertical lines with their names attached to them. The dots on the curves represent their nominal communicability value and the effect of their own removal can be seen as spikes in the other curves.}}
\label{fig:fig3}
 \end{center}
\end{figure}

The results for $I_j$ are depicted in the upper plot of Fig. (\ref{fig:fig3}) for all $473$ characters, where for the sake of clarity only the names of important characters are shown. The removal of most characters basically leaves the entropy unchanged but for Edwin, Oswiu and Paulinus the changes are significant. Among the women, Eanflaed and Ethelburh of Northumbria are the ones who change the topology of the
network more prominently. In this sense these two women are the most relevant ones. Notice that by removing characters one may get a positive or negative variation in entropy value. A positive change means that by removing that particular node the new entropy is smaller, that is the links in the network tend to be less regularly distributed. A negative change means exactly the opposite: the removal of a node makes the entropy larger, links are more evenly distributed: the character removed concentrated around him a larger number of links when compared to the average distribution.

With a list of important characters we re-evaluated the communicability of the four women whose communicability values were more prominent in Fig. (\ref{fig:fig1}): Aelfflaed, Eanflaed, Ethelburh of Northumbria and Hild. This was done by seeing how their communicability values changed when characters were removed one by one. The results are presented in the lower plot of Fig. (\ref{fig:fig3}). The base line of each one fluctuates around their nominal communicability values, that is their values for the full networks. The spikes correspond to their new $C$ values when the characters represented by straight vertical lines were removed. The two women whose values are most affected by the removal of men are Eanflaed (upper blue curve) and Ethelburh of Northumbria (black curve). These changes can be explained due to their kinship to important men: Ethelburh of Northumbria was the wife of King Edwin and mother of Eanflaed, who later married Oswiu. Note also that Ecgfrith's removal -- son of Eanflaed and grandson of Ethelburh -- also affects their communicability, as does the removal of Alhfrith, son of Oswiu through his first wife (Eanflaed was his second).

Less affected by removals are Aelfflaed, daughter of Eanflaed and Oswiu, and Hild. This means that their communicability should be attributed largely to themselves and not to their connection to relevant men. Historically this might be explained by the fact that Aelfflaed was handed over to a religious order at the tender age of one, and was thus most of her life not under the influence of her parents. Her communicability is still mostly affected by the removal of her mother and a bit less by that of her father as one can see in Fig. (\ref{fig:fig3}). Hild, however, who brought Aelfflaed up, was the foundress and abbess of Whitby and  had immense prestige during her lifetime. This is reflected in her 'independent' communicability and these results allow us to get a more nuanced view of their relevance in the network even though their nomimal values of communicability are smaller than that of Eanflaed and Ethelburh of Northumbria. The independent relevance of abbesses in the social networks of this society is a significant finding. While queens may have been more prominent, their power and authority rested on their husbands and was usually lost on the death of the king. Abbesses were better insulated from such vagaries.

\begin{figure}
\begin{center}
\includegraphics[scale=0.5,angle=0]{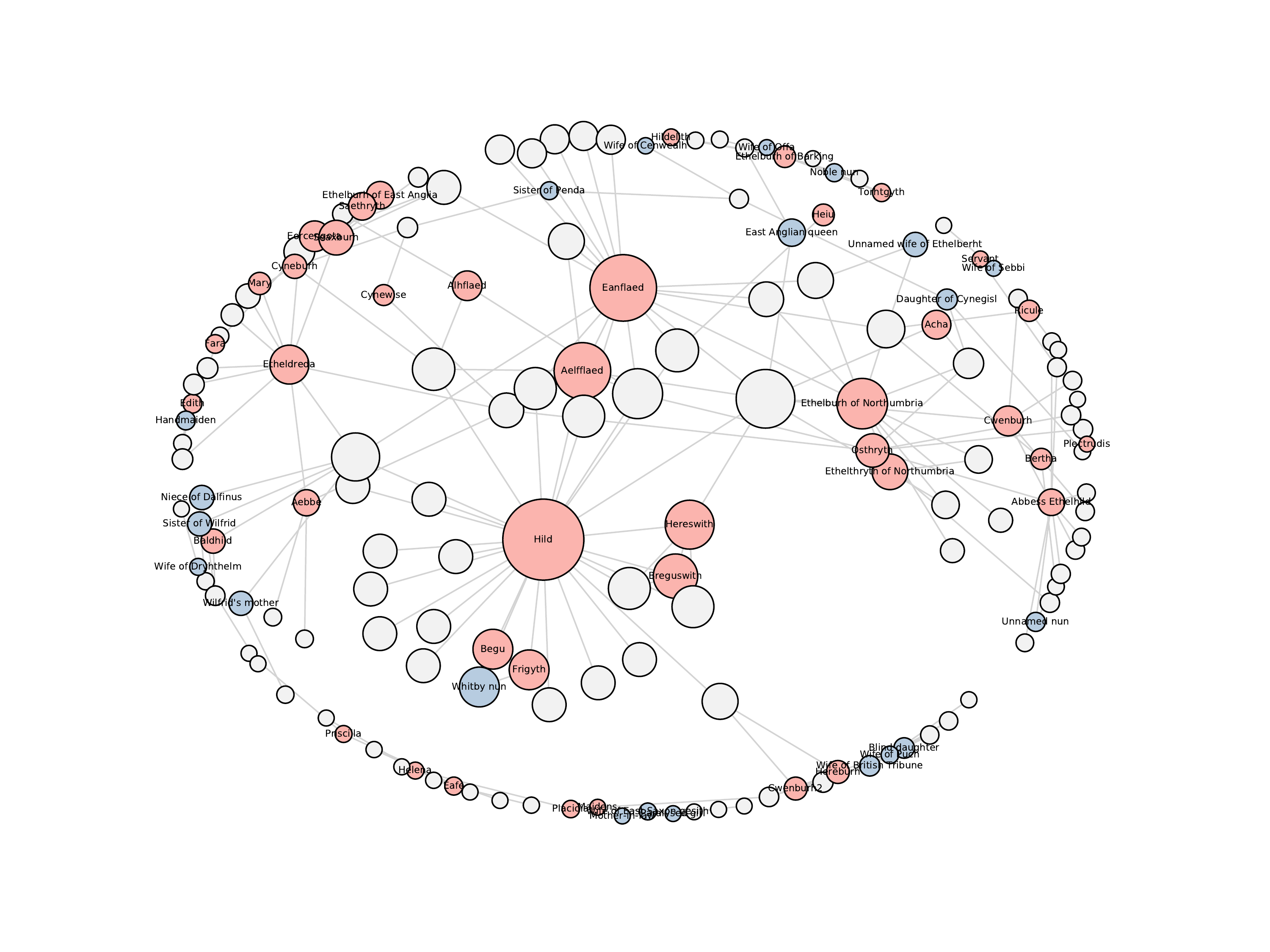}
\includegraphics[scale=0.5,angle=270]{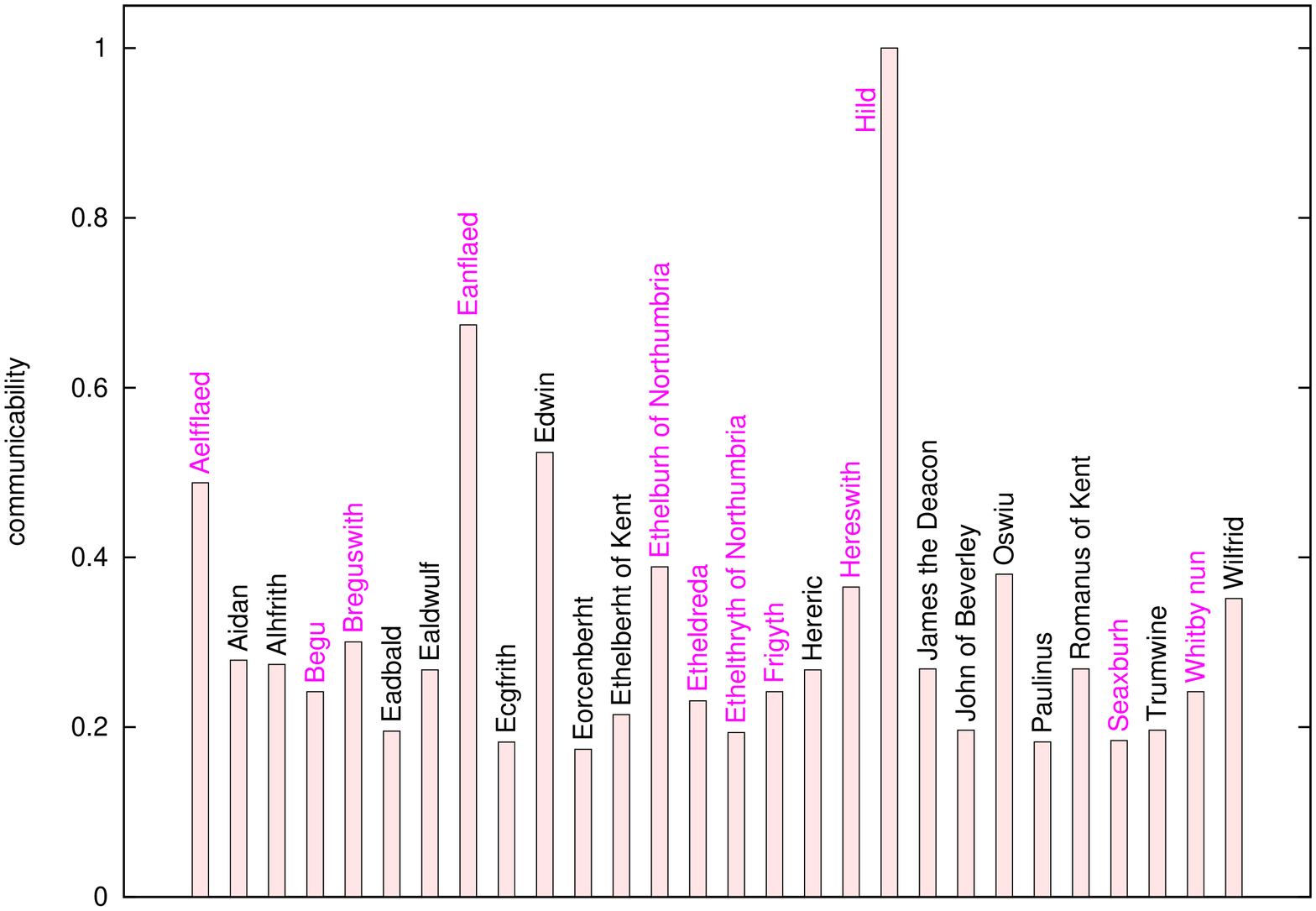}
\caption{{\small The communicability of the women-only- network. The upper figure depicts the network
where the size of the circle is proportional to the communicability of the characters. The lower figure depicts the values
of communicability for the 28 highest-ranking characters. Women's names are magenta, men's black.}}
\label{fig:fig4}
\end{center}
\end{figure}

In order to capture this feature we also studied the communicability of women in Bede's {\it EC} for a 'women-only-network' to see how women compare under themselves. We constructed a women-only-network by keeping men only if they were directly connected to a woman, that is all edges between male characters are not taken into account. One can clearly see from our results, depicted in Fig. (\ref{fig:fig4}), how Hild stands out. In this network, men's communicability is reduced by construction, albeit it is now equal to zero as some still represent nodes in the paths of women.

\begin{figure}
\begin{center}
\includegraphics[scale=0.25,angle=0]{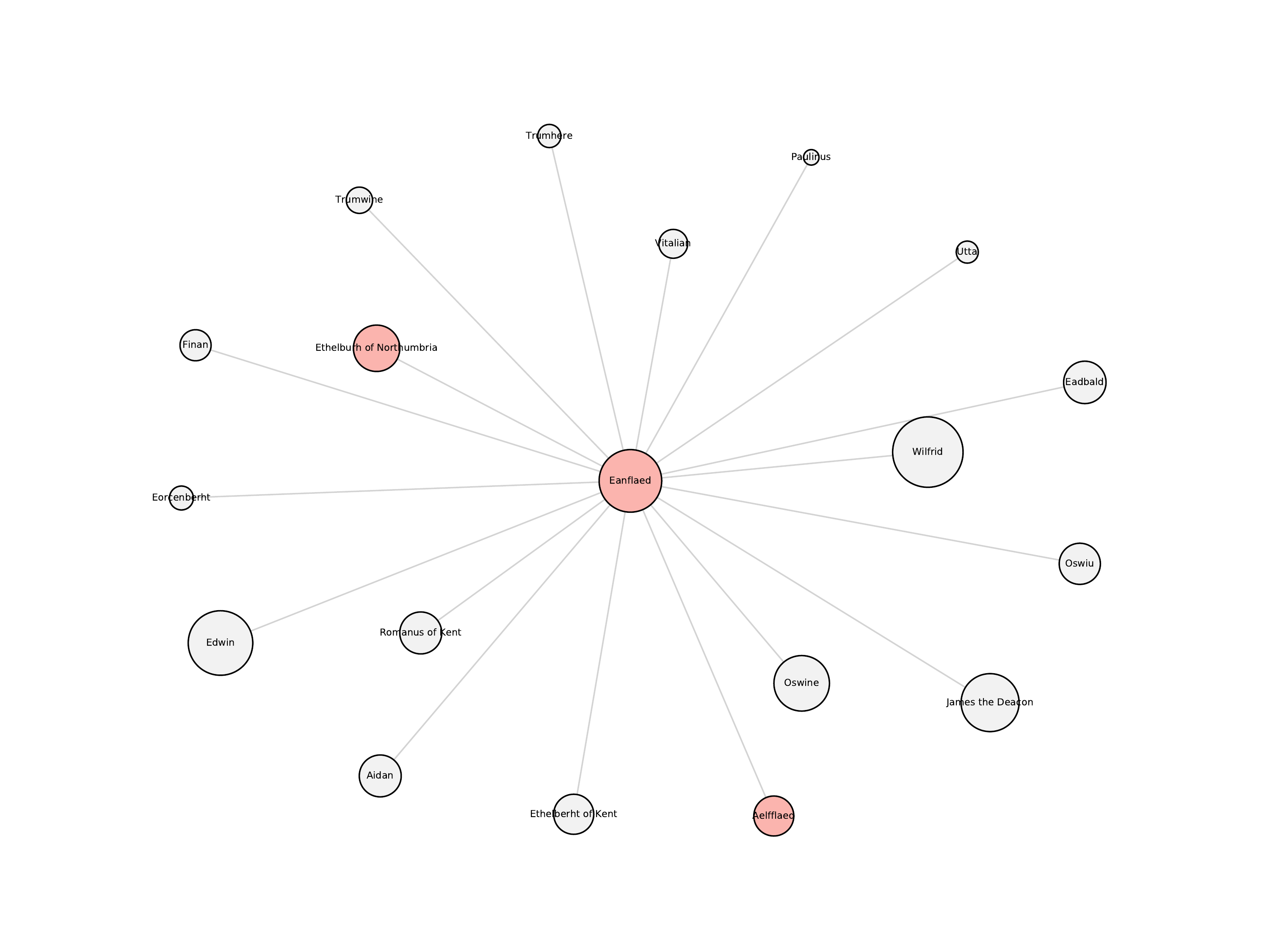}
\includegraphics[scale=0.25,angle=0]{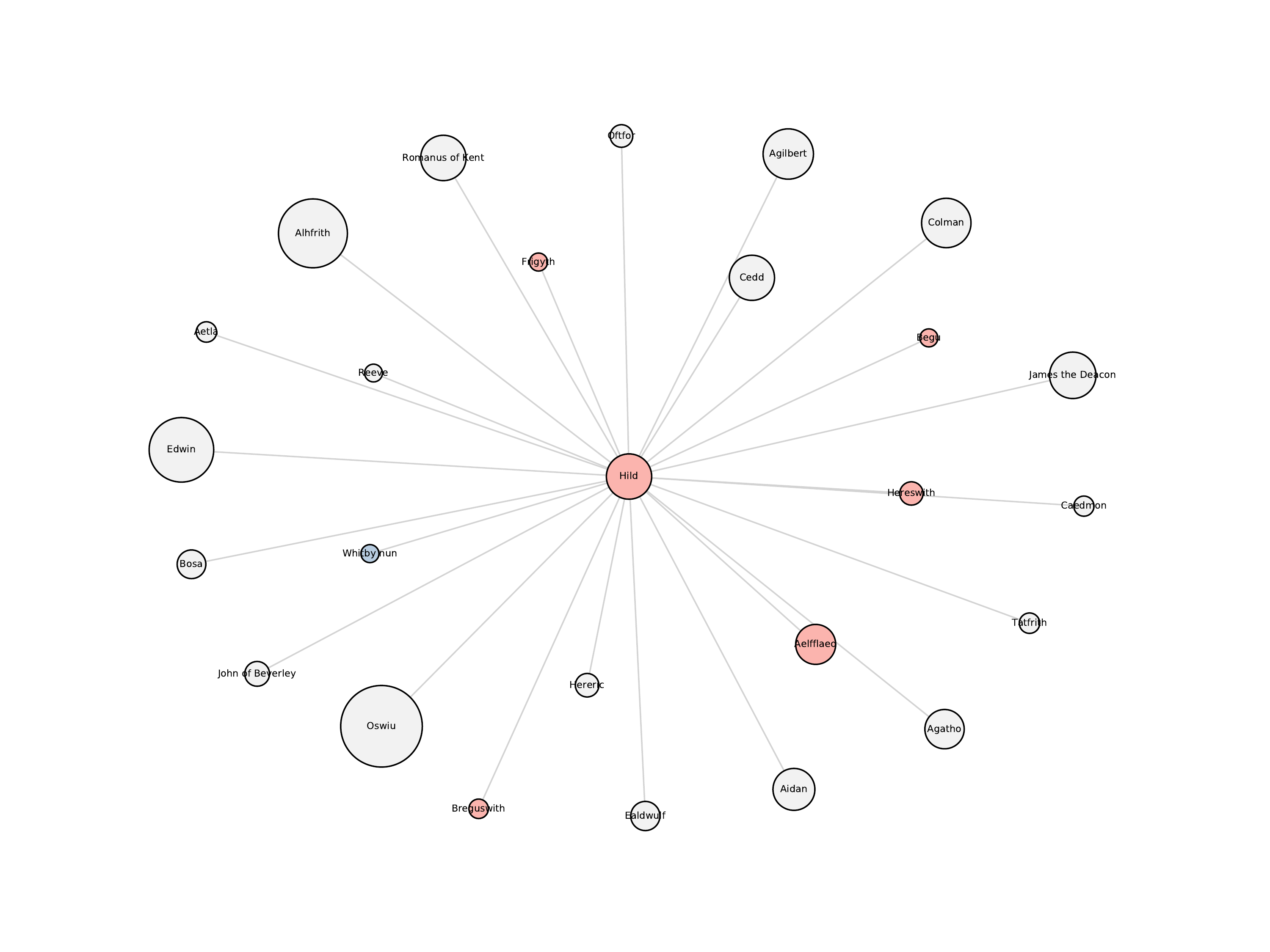} \\
\includegraphics[scale=0.25,angle=0]{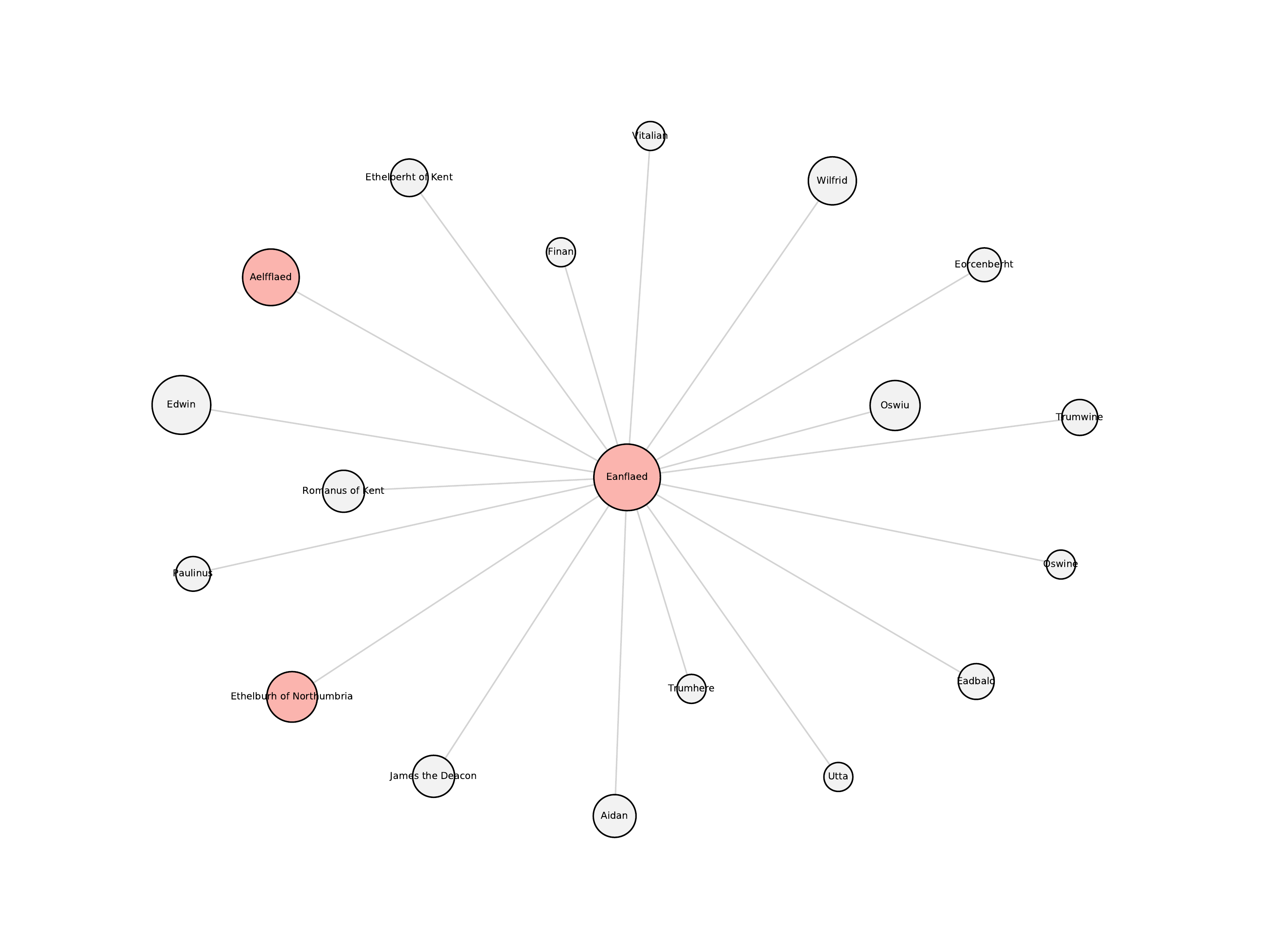}
\includegraphics[scale=0.25,angle=0]{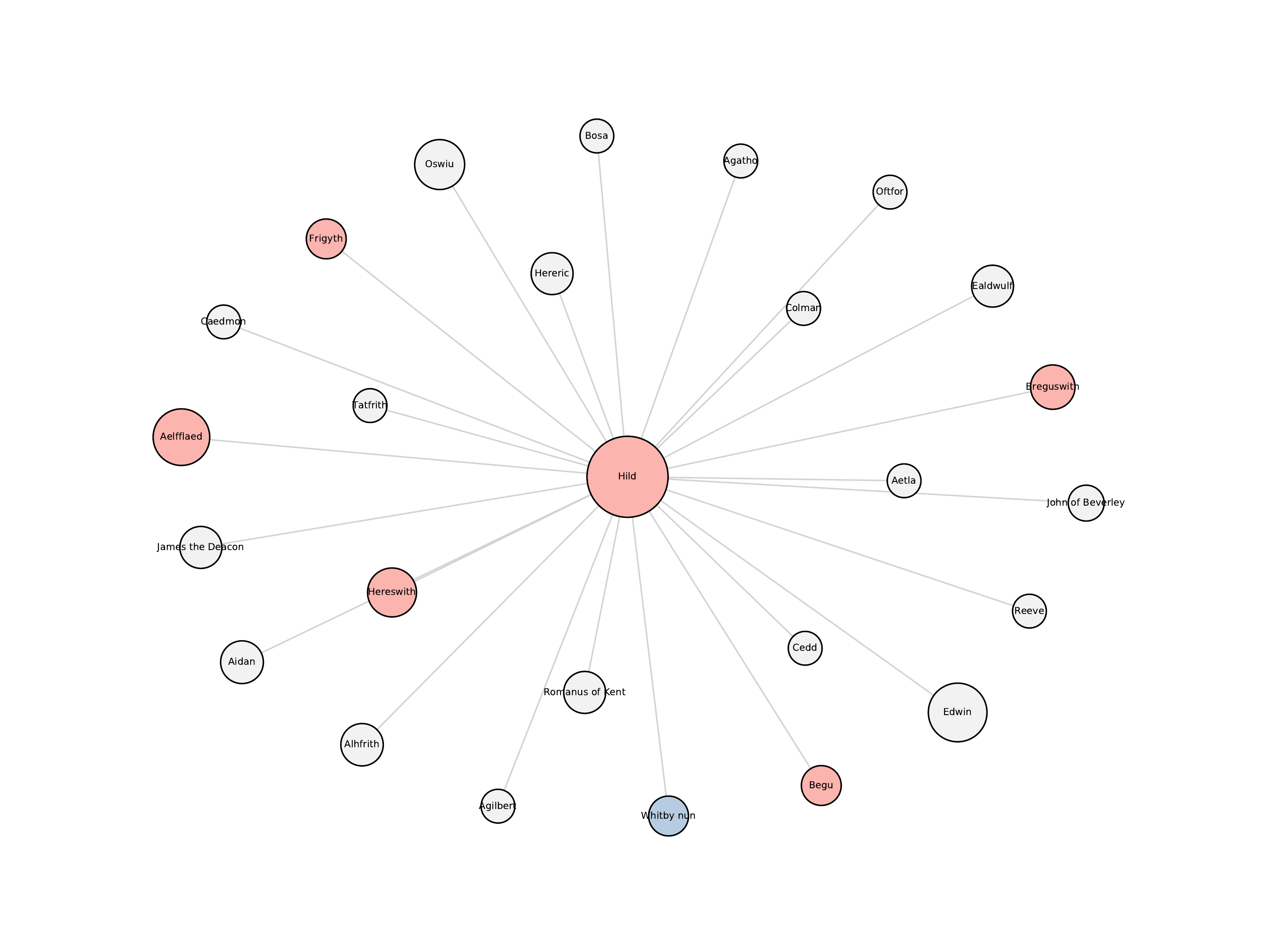} 
\caption{{\small Eanflaed and Hild's communicability values -- represented as networks -- for the complete network (upper graphs) and the women-only-network (lower graphs).}}
\label{fig:fig5}
\end{center}
\end{figure}

The contrast between the whole network and the women-only-network and how if affects Hild and Eanflaed's communicability can be also seen in Fig. (\ref{fig:fig5}). In this figure we plot only Hild and Eanflaed's direct connections. The upper two figures represent their communicability in the complete network, where one clearly sees that Eanflaed's communicability is higher (larger central circle). However, when we represent the women-only-network in the same way, Hild's communicability value surpasses that of Eanflaed, showing clearly how in the former case Eanflaed's score is due mainly through her connections with important men. Another relevant feature of these plots is that women are sparsely connected to each other, although Hild is the one with the largest number of connections: 5.

Any quantitative method of finding key players or the most influential character or set of characters in a representation of a historical network is highly contingent upon the way we look at them or the way we weigh and classify their connections. This fact is not new and is widely stated in the literature of social networks \cite{Borgatti2006,Arroyo2010}.

In table 2 we show three usual centrality measures for the six women with highest degrees: normalized degree, eigenvector centrality, and betweenness.

\begin{center}
\begin{tabular}{lccccc}
\toprule
 women && degree & eigenvector centrality & betweenness \\
\midrule
Eanflaed  && 0.127 & 0.251    & 0.263  \\ 
Hild      && 0.119 & 0.119    & 0.183  \\ 
Ethelburh of Northumbria && 0.081 & 0.120 & 0.024 \\ 
Etheldreda && 0.051 & 0.038   & 0.150  \\ 
Seaxburh && 0.041& 0.077   & 0.037 \\ 
Aelfflaed    && 0.031& 0.098   & 0.089  \\
\bottomrule
\end{tabular}

\label{tab:table2}
\end{center}
\vskip 0.5cm
{\small Table 2: Degree, eigenvector centrality and betweenness (all normalized) for the six highest-ranking women in the aggregate network.}

Eanflaed is the woman with the highest scores (Table 2), and the one with the highest communicability (Fig. \ref{fig:fig1}). On the other hand, her communicability is more susceptible to the presence of prominent men than Hild (see Fig. (\ref{fig:fig3})). Hild's centrality values place her as the second highest ranking woman, even though her eigenvector centrality is half of that of Eanflaed. Also, Aelfflaed's communicability is larger than the East Anglian princesses, Etheldreda and Seaxburh (both not shown in Fig. (\ref{fig:fig3})), since she has higher eigenvector centrality than both of them and smaller betweenness when compared to Seaxburh. All leads us to conclude that communicability seems to combine the information one obtains from the eigenvector centrality with that from betweenness. When studied in association with the entropy measure of relevance, it also reveals which characters are unaffected by the removal of high-ranking characters and therefore can be seen as being relevant on their own merit.

\section{Conclusions}

A conventional historical analysis of Bede's {\it HE} prioritises the actions and interactions of male characters. Some historians have speculated that the women described by Bede had power, but the current consensus is that women at the time were only powerful through their connections with men~\cite{Nicholson1978,Hollis1992,Wade2013}.

In this paper we have used communicability which quantifies the propensity of communication between two characters and the notion of relevance taken from the topological variation caused by the removal of a character (entropy variation) as a way to quantify what we could call their numerical importance. Our approach has shown that it is true that some women's power, for example that of queens, was dependent upon their male connections, such as husbands, sons and even step-sons. However, we have identified at least one woman, Hild, abbess of Whitby, whose position in the network as derived from Bede was unaffected by her male connections. We are applying this approach to other texts from the period to corroborate our findings.

Our computation by no means replaces other methods of assessing the historical relevance of these women. However, the idea of applying tools of social networks to historical accounts is an attempt to bring to the fore different interpretations or less intuitive views on the role of these women.  

\section*{Acknowledgments}
The authors would like to acknowledge the financial support of the Leverhulme Trust Grant RPG 2018--014 'Women, Conflict and Peace: Gendered Networks in Early Medieval Narratives'. ALCB, SRD and SDP would like to thank the University of Sheffield and University College Cork for the hospitality. MMC and JH would like to
thanks the Universidade Federal de Porto Alegre for the hospitality.

\end{document}